\documentclass[epsf]{aa}  

\def\simlt{\lower.5ex\hbox{$\; \buildrel < \over \sim \;$}}
\def\simgt{\lower.5ex\hbox{$\; \buildrel > \over \sim \;$}} 
\def\lmcx3{LMC~X-3} 
\def\xmm{\it XMM-Newton} 
\def\xte{\it RXTE}

\begin{document}

\thesaurus{01        
          (02.01.2;  
	   02.02.1;  
	   08.02.3;
           08.05.1;  
	   08.23.3;  
           13.25.3)} 

\title{XMM-Newton EPIC and RGS observations of LMC X-3} 


\titlerunning{EPIC and RGS observations of LMC X-3}

\author{K. Wu \inst{1}\fnmsep\inst{2},   
        R. Soria\inst{1}, 
        M. J. Page\inst{1},          
        I. Sakelliou\inst{1},  
        S. M.\ Kahn\inst{3} 
        \and C. P. de Vries\inst{4} } 
 
\authorrunning{K. Wu et al.}

\offprints{K. Wu}

\institute{Mullard Space Science Laboratory, 
           University College London,
           Holmbury St Mary, Dorking, RH5 6NT, UK
           \and
           Research Centre for Theoretical Astrophysics, 
           School of Physics, 
           University of Sydney, NSW 2006, Australia                 
           \and 
           Columbia Astrophysics Laboratory, 
           Columbia University, 
           550 West 120th Street, New York, NY 10027, USA 
           \and 
           Space Research Organization of the Netherlands, 
           Sorbonnelaan 2, 3584 CA Utrecht, The Netherlands      
             }

\date{
      }

\maketitle

\begin{abstract}
         We report the results of preliminary analysis 
           of the {\xmm} EPIC and RGS observations      
           of the candidate black-hole binary \lmcx3 
           between February and June 2000.  
         The observations covered 
           both the soft and the hard X-ray spectral states.  
         The hard-state spectra were dominated by a power-law component      
           with a photon index $\Gamma \approx1.9$. 
         The soft-state spectra consisted of a thermal component  
           with a multi-colour disk temperature $T_{\rm in}$ 
                of $\sim 0.9$~keV   
           and a power-law tail with $\Gamma \approx 2.7$.  
         The model in which     
           the X-rays from \lmcx3 in the high-soft state   
           are powered by a strong stellar wind 
           from a massive companion 
           is not supported by  
           the small line-of-sight absorption 
           ($n_{\rm H} \simlt  10^{21}$~cm$^{-2}$)  
           deduced from the RGS data.    
         The transition from the soft to the hard state     
           appears to be a continuous process 
           associated with the changes in the mass-transfer rate.     
          
\end{abstract}

  
\section{Introduction}

\lmcx3 (Leong et al.\ 1971) is a persistent X-ray source in the 
  Large Magellanic Cloud (LMC).  
It has an orbital period of 1.70~d,  
  and a mass function $\simeq 2.3 M_{\odot}$ (Cowley et al.\ 1983).  
The optical brightness of  $V \sim 17$ 
  indicates that the system has a massive companion 
  (van Paradijs et al.\ 1987) . 
The companion is often classified as a B3\,V star 
  (Cowley et al.\ 1983), although a B5\,IV spectral type has also been 
  suggested (Soria et al.\ 2001).     
The non-detection of eclipses in the X-ray curve    
  implies that the orbital inclination of the system 
  is \simlt~$70^{\circ}$ (Cowley et al.\ 1983).  
The inferred mass of the compact star is \simgt~$7 M_{\odot}$ 
  (Paczynski 1983),   
  thus establishing that the system is a black-hole candidate (BHC). 
 
Most BHCs show soft and hard X-ray spectral states.   
Their X-ray spectrum in the soft state 
  generally consists of both a thermal and a power-law component. 
The thermal component can be fitted by a blackbody spectrum  
  with a temperature $\sim 1$~keV, and      
  it is interpreted as thermal emission from the inner accretion disk.   
The power-law component is believed to be comptonised emission 
  from a disk corona (Sunyaev \& Titarchuk 1980) 
  or from the high-speed infalling plasma  
  near the black-hole event horizon (Titarchuk \& Zannias 1998). 
The photon index $\Gamma$ of the power law is $ \approx 2.5$--4.  
In the hard state, the thermal component is insignificant.  
The spectrum is dominated 
  by a flat ($1.5 \simlt \Gamma \simlt 2$), extended power-law component.   

\lmcx3 is found in the soft state most of the time.   
LMC X-1, another high-mass BHC in the LMC, 
  has been seen in the soft state only.   
In contrast, Cyg X-1, the high-mass BHC in our Galaxy, 
  tends to be in the hard state for the majority of the time. 
Such preference for either the hard or the soft state 
  is not evident in the low-mass transient BHCs 
  (e.g.\ GRO~J1655$-$40).  
  It is still unclear how mass transfer occurs in \lmcx3:  
  whether via Roche-lobe overflow or capture of the stellar 
  wind from the companion.  
Given the fact that a B3\,V star is unable 
  to drive a strong stellar wind 
  or to fill its Roche lobe if it is in a binary system 
  with a 1.7-d orbital period,       
  it is difficult to explain the observed X-ray luminosity of the system. 

In this letter, we report on the {\xmm} observations of \lmcx3  
  between 2000 February and June.    
The system was in an unusual hard state in April, 
  and our observations span the transition from 
  the soft state to the hard state and then back to a softer state. 
Making use of these new X-ray spectral data, 
  we attempt to shed light on the question 
  of mass transfer in the system (see also Soria et al.\ 2001).  
    
\begin{table}
\begin{minipage}{80mm}
\caption{{\xmm} EPIC Observation Log}
\label{mathmode} 
\centering 
\begin{tabular}{@{}lllll} 
\hline
  Revolution/ID &  Instrument
  & Exposure
   \footnote{For the PN small window mode, the live time was 71\% 
of the exposure time listed here; for the MOS RFS mode the live time 
was 6.9\% of the exposure time.}   
  & Mode 
   \footnote{See {\it XMM} Remote Proposal 
    Submission Software Users' Manual}  
  \\ \hline \hline 
 && \\
  Rev0028/201 &  PN    
                       & 08.1~ks    & small window \\[2pt]             
  Rev0028/301 &  PN  
                       & 0.3~ks    & full PN      \\[2pt] 
          &  MOS1   
                       & 08.9~ks    & partial RFS  \\[2pt] 
  Rev0030/501 & MOS1  
                       & 05.9~ks    & partial W5   \\[2pt] 
  Rev0041/101 &  PN     
                       & 0.7~ks    & full PN       \\[2pt]
  Rev0041/401 & MOS1  
                       & 03.5~ks    & partial W2   \\[2pt] 
          &  MOS1   
                       & 03.6~ks    & partial W5   \\[2pt]             
          &  MOS1   
                       & 03.4~ks    & partial W4   \\[2pt]   
  Rev0045/101 &  PN    
                       & 01.6~ks    & full PN      \\[2pt]             
  Rev0045/301 & MOS1  
                       & 06.6~ks    & partial W3   \\[2pt]    
  Rev0066/101 &  PN    
                       & 26.6~ks 
                       \footnote{Only the first 8.3~ks are useful}    
                                    & small window \\[2pt]    
  Rev0092/201 &  PN     
                       & 24.2~ks    & small window \\[2pt]        
            &  MOS1  
                       & 20.3~ks    & partial W4 \\[2pt] 
\hline    
\end{tabular}   
\end{minipage} 
\end{table}  

\begin{table}
\begin{minipage}{80mm}
\caption{{\xmm} RGS Observation Log }
\label{mathmode} 
\centering 
\begin{tabular}{@{}lllll} 
\hline
   Revolution/ID & Instrument \footnote{The labels ``s\#'' identifies 
   multiple exposures during the same revolution; for more details  
   see the Preferred Observation Sequence 
   files of each revolution.}
  & Exposure 
  & count rate \footnote{In photons~s$^{-1}$} 
  \\ \hline \hline 
 && \\
  Rev0028/201 & RGS2 (s1) 
                       & 17.6~ks & 4.79$\pm$0.02 \\[2pt] 
  Rev0030/501 & RGS2 (tot) 
                       & 12.1~ks & 4.52$\pm$0.03 \\[2pt]
   Rev0045/101             & RGS1 (s1) 
                       & 06.4~ks & 3.59$\pm$0.03 \\[2pt]  
& RGS2 (s1) 
                       & 06.4~ks & 3.56$\pm$0.03 \\[2pt]            
  Rev0045/201 & RGS1 (s1) 
                       & 06.7~ks & 3.62$\pm$0.02 \\[2pt] 
            & RGS2 (s1) 
                       & 06.8~ks & 3.57$\pm$0.02 \\[2pt] 
  Rev0045/301 & RGS1 (s1) 
                       & 03.0~ks & 3.66$\pm$0.04 \\[2pt] 
            & RGS1 (s3) 
                       & 03.0~ks & 3.57$\pm$0.04 \\[2pt]   
            & RGS1 (s5) 
                       & 03.0~ks & 3.40$\pm$0.04 \\[2pt]     
            & RGS2 (s2) 
                       & 00.3~ks & 3.63$\pm$0.17 \\[2pt]  
            & RGS2 (s4) 
                       & 03.0~ks & 3.26$\pm$0.04 \\[2pt]   
            & RGS2 (s6) 
                       & 00.3~ks & 3.18$\pm$0.17 \\[2pt]   
  Rev0066/101             & RGS1 (s7) 
                       & 44.3~ks & 0.0154$\pm$0.0012 \\[2pt]  
& RGS2 (s8) 
                       & 44.3~ks & 0.0153$\pm$0.0012 \\[2pt]  
  Rev0092/101             & RGS1 (s2) 
                       & 51.9~ks & 1.32$\pm$0.01 \\[2pt]  
& RGS2 (s2) 
                       & 51.9~ks & 1.53$\pm$0.01 \\[2pt]            
  Rev0092/201 & RGS1 (s4) 
                       & 25.6~ks & 1.56$\pm$0.01 \\[2pt] 
            & RGS2 (s5) 
                       & 25.6~ks & 1.34$\pm$0.01 \\[2pt]           
\hline   
\end{tabular}  
\end{minipage} 
\end{table}

\begin{figure}
\begin{center}
\setlength{\unitlength}{1cm}
\begin{picture}(8.8,13.5)
\put(0,9){\includegraphics{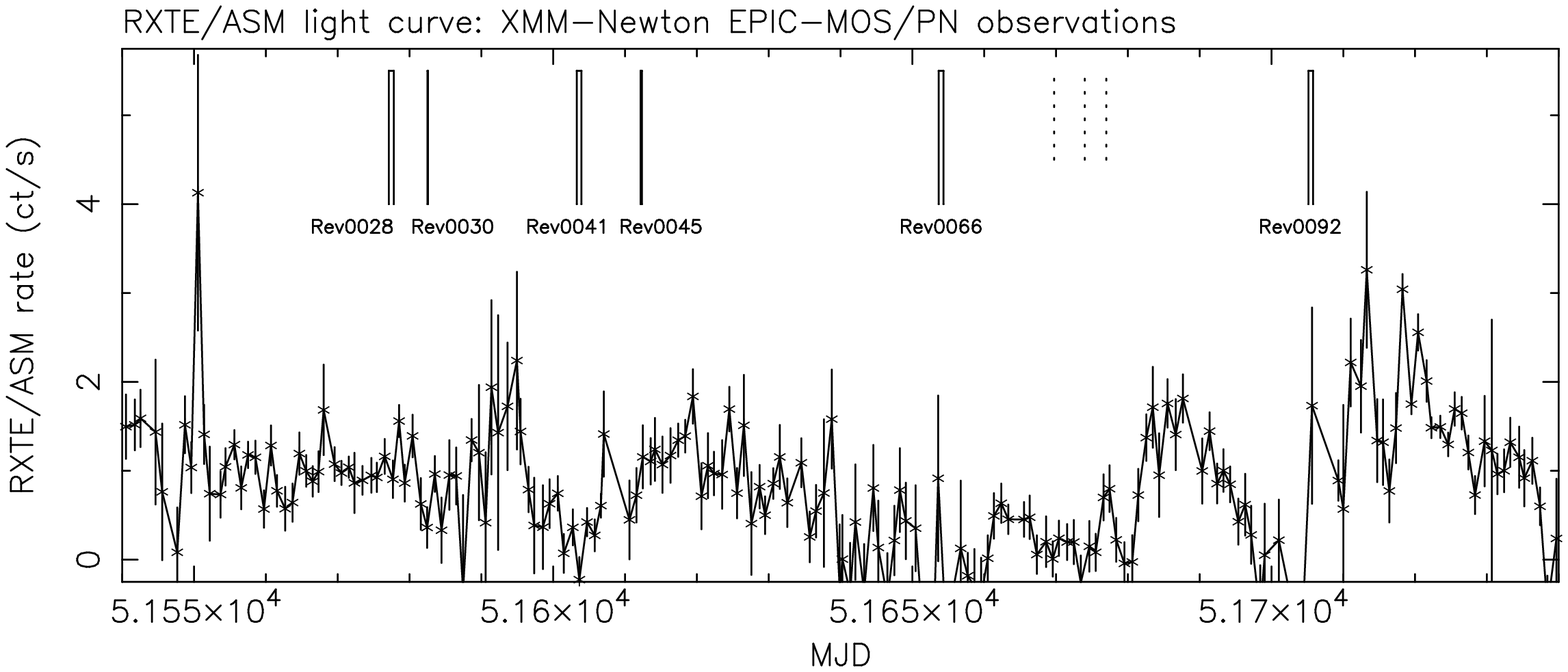} } 
\put(0,6){\includegraphics{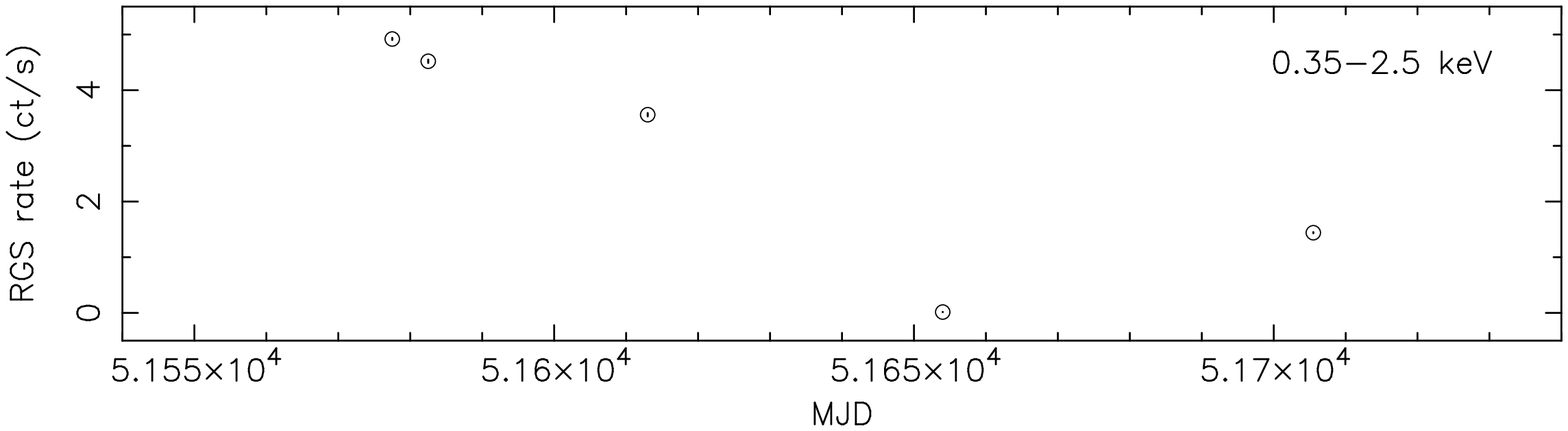} }   
\put(0,3){\includegraphics{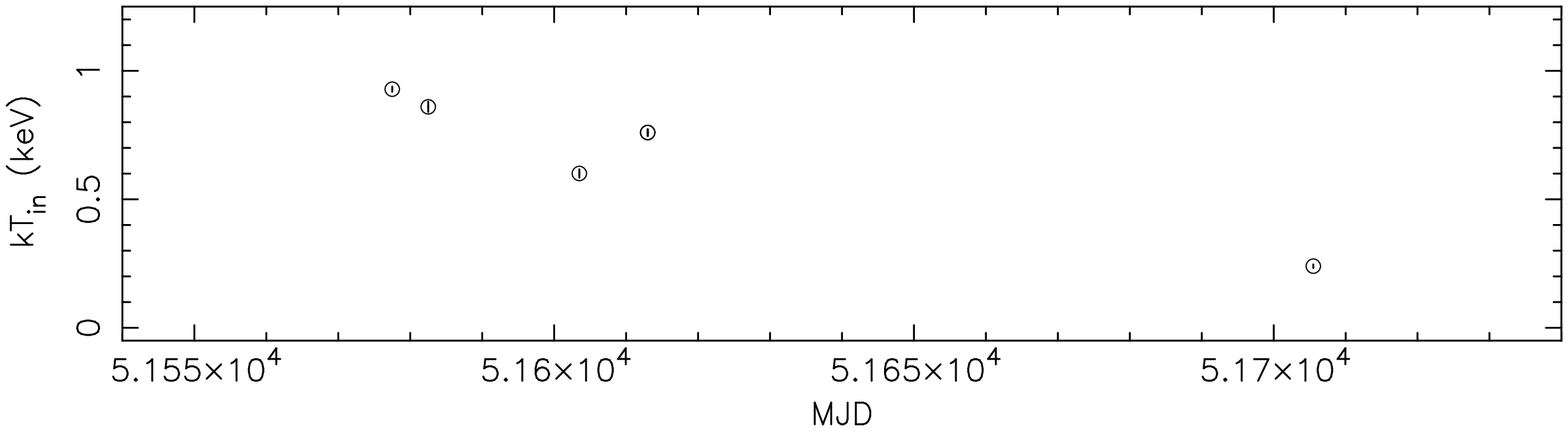} } 
\put(0,0){\includegraphics{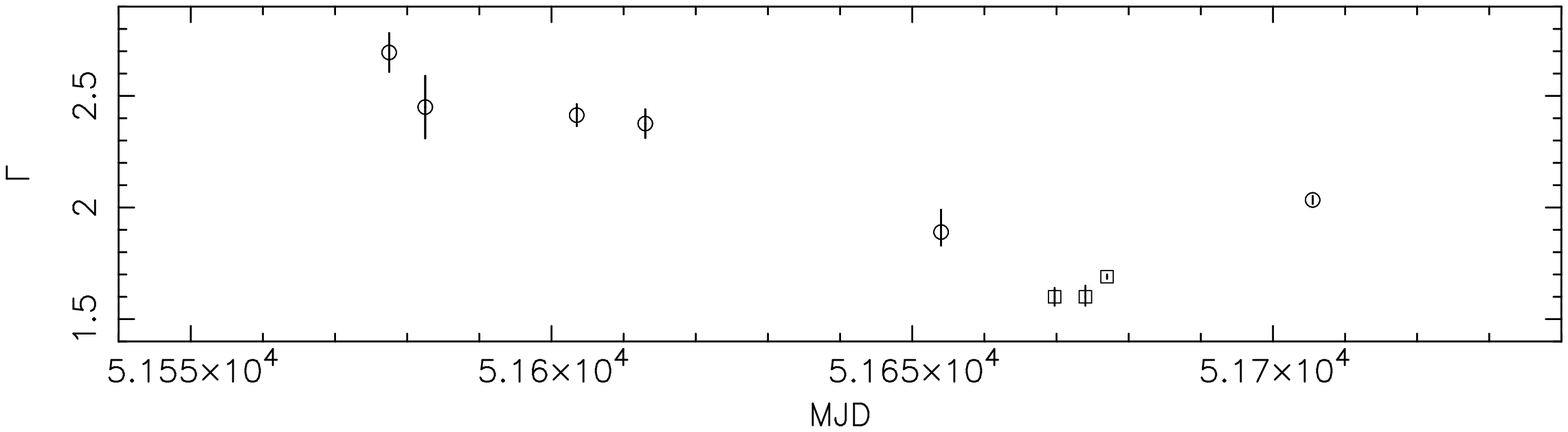} } 
\end{picture}
\caption{
   (First panel from the top): The {\xmm} EPIC observations of \lmcx3, 
            marked by solid vertical lines,            
            are shown and compared 
            with the {\xte}/ASM light curve. 
          The three {\xte}/PCA observations 
            that confirmed the hard state of the system 
            (Boyd et al.\  2000) are marked by dotted vertical lines.  
   (Second panel): The 0.35$-$2.5~keV light curve 
             from the RGS observations. 
   (Third panel): The evolution of the fit temperature 
            of the thermal component in the EPIC-PN and MOS spectra.   
          The open circles  
             are the means of the fit disk-blackbody temperatures 
             $T_{\rm in}$ error-weighted  
             for the observations during the same revolution. 
   (Fourth panel): The evolution of the photon index $\Gamma$ 
             of the power-law tail.   
          The data points (open circles) are error-weighted means.  
          The open squares are 
             the photon indices obtained by Boyd et al.\ (2000)  
             from the three {\xte}/PCA observations on May 5.76, 10.01 and 
             13.94~UT; 
             a model with a single power law was used.  }
\label{fig1}  
\end{center} 
\end{figure}   

\section{Observations} 

\subsection{Technical details}

\lmcx3 was observed with 
  the European Photon Imaging Camera (EPIC; Turner et al.\ 2001), 
  the Reflection Grating Spectrograph (RGS; Brinkman et al.\ 2001) and  
  the Optical Monitor (OM; Mason et al.\ 2001) on board {\xmm}, 
  during the Performance Verification Phase  
  between 2000 February and June.  
Here we present part of the EPIC and RGS observations.    
The OM observations are presented in Soria et al.\ (2001).  
The logs of the EPIC-MOS1/PN and RGS observations used for this study 
  are shown in Tables 1 and 2 respectively.    
The EPIC-MOS2 observations are not presented, 
  as a reliable response matrix is not yet available.    
The exposure times listed in Table 1 
  take into account the counting mode interruptions. 
In some cases, the fraction of live time 
  during which photons were actually collected 
  is significantly smaller than the total exposure time.   

All EPIC exposures were taken with the ``medium'' filter;   
Some of the data suffer from pile-up 
   when the source was observed in the bright state (before Rev0066). 
MOS exposures taken in the ``full window'' mode are worst affected, 
   and so are excluded from the analysis. 
``Partial window'' MOS exposures and ``full window'' PN exposures 
  are, however, less affected. 
In our analysis, we have removed the central pixels 
  (which are most affected by pile-up) 
  from the extraction regions, 
  for the partial MOS and full PN exposures that are affected. 
The corresponding normalisations of the fitted spectral models 
  for these observations are therefore smaller than the true values,
  and they are put in brackets in Table 3. 
We are aware that 
  the photon index of the power-law component 
  is affected by pile-up. 
For the Rev0028, 0030, 0041 and 0045 observations, 
  we estimate that the fitted values 
  are $\approx 0.15$--$0.20$ smaller (harder) than the real value.
The ``small window'' PN exposures 
  are not significantly affected by pile up, 
  and they are more reliable 
  for the determination of the power-law photon indices.

The data were processed using the September 2000 release 
of the SAS, except for the PN small window mode data, 
which were processed with the October 2000 SAS release. 
All spectral fits presented here were performed with 
the most up-to-date response matrices available 
at the end of October 2000 (mos1\_medium\_all\_qe17\_rmf3\_tel5\_15.rsp 
for EPIC-MOS1 and epn\_fs20\_sY9\_medium.rmf for EPIC-PN). 
We are aware that the PN response matrix assumes 
a nominal event threshold of 20 adu, whereas a threshold 
of 23 adu was used in the Rev0028 and Rev0041 observations. 
This may have a small effect on the low-energy end of the spectrum 
and make the determination of the column density from the early PN data 
less reliable. We used the RGS spectra to obtain a better 
determination of the column density. 

\subsection{The 2000 April low-hard state}
   
The X-ray luminosity of \lmcx3 appeared to be declining    
  during our February--March observations, 
  with the {\xte}/ASM $2-20$~keV count rate 
  generally below 3~ct~s$^{-1}$ (Figure~1).    
The Rev0066/101 observation (2000 April 19) was carried out 
  around the middle of a faint-hard state, 
  where the {\xte}/ASM count rate was consistent with zero.   
The {\xte}/PCA data obtained on May 5.76 and 10.01~UT 
 showed power-law spectra with a photon index $1.60\pm 0.05$  
  and a soft ($2$--$10$~keV) X-ray flux 
  of $\approx 5$--$9\times 10^{36}~{\rm erg}~{\rm s}^{-1}$ 
  at 50~kpc (Boyd \& Smale 2000; Boyd et al.\ 2000).   
The system seemed to be in the process 
  of returning to the soft state 
  in the Rev0092 (June 9--10) observations.   

\subsection{EPIC MOS and PN data}  

We consider a conventional model 
   consisting of an absorbed multi-colour disk blackbody
        plus a power law, 
        wabs*zvfeabs*(diskbb+powerlaw) in XSPEC, 
   to fit the EPIC data.  
As the fitting process is limited
    by the reliability of the response matrices currently available,
    more sophisticated models or the inclusion of additional features
    (such as lines) are not warranted.  
The generic models used here are therefore sufficient
    for this preliminary analysis.  

For the MOS observations, we consider only the channels 
   in the $0.3$--$8.0$~keV energy range; 
   for the PN observations we extend the energy range 
   to $0.3$--$12.0$~keV.
The data are binned with the {\small{FTOOLS}} task grppha,
   such that the number of counts in each group of channels 
   is larger than 40. 
This improves the signal-to-noise ratio
   for energies $\simgt 6$~keV, 
   while it does not affect low-energy channels. 
We have also checked that different
   binning criteria give consistent results.
No systematic error has been added to the data. 

We consider the line-of-sight Galactic absorption toward 
   LMC X-3 to be $3.2\times 10^{20}$~cm$^{-2}$ 
   following Wilms et al.\ (2000), and 
   use it as the value for the (fixed) first absorption component.
The second absorption component is then 
   the intrinsic photoelectric absorption 
   within the binary system and the LMC.
We also fix both the iron abundance and the total metal abundance 
   to be 0.4 times the solar values, to account for the 
   lower metallicity of the LMC (Caputo et al.\ 1999).
The best fit parameters are shown in Table 3. 
Figures 2, 3 and 4 show the EPIC-PN spectra 
  from the Rev0028/201, Rev0066/101 and Rev0092/201 observations 
  respectively (all taken in small window mode, not affected 
  by pile-up).      

The reduced $\chi_\nu^2$ are about one in most fits. 
The large $\chi_\nu^2$ in some fits 
  (e.g. the MOS1 observations in Rev0092/201) 
  are probably due 
  to the uncertain response and effective-area calibrations. 
In particular, the feature seen at $\approx 0.5$ keV 
  in the PN spectra (Figures 2, 3, 4) is due to uncertainties 
  in the charge transfer efficiency correction for the small window mode.  
Values of $\chi_\nu^2 \le 1.05$ are obtained 
  when the $0.4$--$0.6$ keV energy range is excluded from the fit.

\begin{table*}
\begin{minipage}{150mm}
\caption{Fits to the EPIC data: absorbed disk blackbody and power law} 
\label{mathmode} 
\centering 
\begin{tabular}{@{}lllccccclcc} 
\hline
 & Observation  &  & $n_{\rm H}$\footnote{ 
   The Galactic line-of-sight absorption  
       ($= 3.2\times 10^{20}$~cm$^{-2}$) has been subtracted.
   We assumed a metallicity of 0.4 times the solar value, i.e., Z = 0.008.}  
       &  $T_{\rm in}$  & $A_{\rm disk}$ &  $\Gamma$  
       & $A_{\rm pl}$ & $\chi_\nu^2$(dof)      
  \\ \hline \hline 
 &&&&&&&&\\
 & Rev0028/201 & PN     & $15.65^{+2.10}_{-1.96}\times 10^{20}$  
            & $0.93^{+0.01}_{-0.01}$ & $4.32^{+0.24}_{-0.25}$ 
            & $2.71^{+0.09}_{-0.08}$ 
                  & $8.92^{+0.83}_{-0.79}\times 10^{-3}$  
            & 1.31 (784) \\[2pt]  
 & Rev0028/301 & PN & $4.82^{+9.34}_{-4.82}\times 10^{20}$ 
            & $0.93^{+0.04}_{-0.05}$ & $(6.26^{+1.38}_{-1.15})$   
            & $2.45^{+0.53}_{-0.44}$ 
                   & $(6.04^{+4.13}_{-3.58} \times 10^{-3})$ 
            & 1.02 (192) \\[2pt] 
           &  & MOS1 (s7) & $17.98^{+8.62}_{-8.12} \times 10^{20}$ 
            & $0.88^{+0.06}_{-0.06}$ & $(0.65^{+0.24}_{-0.15})$   
              & $2.53^{+0.41}_{-0.35}$ 
                   & $(1.55^{+0.46}_{-0.50} \times 10^{-3})$ 
            & 0.75 (170)  \\[2pt]  
 & Rev0030/501 & MOS1 (s7) & $13.00^{+3.48}_{-3.53}\times 10^{20}$ 
            & $0.86^{+0.02}_{-0.02}$ & $(6.10^{+0.74}_{-0.65}$  
            & $2.45^{+0.14}_{-0.14}$ 
                   & $(9.60^{+1.65}_{-1.70} \times 10^{-3})$ 
            & 1.28 (282) \\[2pt]  
 & Rev0041/101  & PN & $13.66^{+6.17}_{-6.94}\times 10^{20}$ 
            & $0.59^{+0.07}_{-0.06}$ & $(10.91^{+4.13}_{-5.17})$   
            & $2.57^{+0.22}_{-0.27}$ 
                   & $(10.60^{+3.05}_{-3.43} \times 10^{-3})$ 
            & 1.01 (259) \\[2pt] 
 & Rev0041/401  & MOS1 (s3) & $15.08^{+2.59}_{-2.69}\times 10^{20}$ 
            & $0.62^{+0.04}_{-0.03}$ & $16.80^{+5.25}_{-4.36}$    
            & $2.47^{+0.08}_{-0.08}$ 
                   & $2.97^{+0.32}_{-0.33}\times 10^{-2}$ 
            & 1.12 (260) \\[2pt] 
           &  & MOS1 (s5) & $10.31^{+2.56}_{-2.63}\times 10^{20}$ 
            & $0.64^{+0.03}_{-0.02}$ & $16.65^{+3.85}_{-3.37}$   
            & $2.33^{+0.08}_{-0.09}$ 
                    & $2.20^{+0.27}_{-0.28} \times 10^{-2}$ 
            & 1.22 (258)  \\[2pt] 
           &  & MOS1 (s7) & $13.28^{+2.75}_{-2.80}\times 10^{20}$ 
            & $0.58^{+0.02}_{-0.02}$ & $30.38^{+6.62}_{-6.12}$   
            & $2.41^{+0.09}_{-0.09}$ 
                    & $2.80^{+0.37}_{-0.38} \times 10^{-2}$ 
            & 1.25 (258)  \\[2pt] 
 & Rev0045/101 & PN & $6.15^{+3.65}_{-3.65}\times 10^{20}$ 
            & $0.76^{+0.03}_{-0.02}$ & $(9.00^{+1.46}_{-1.33})$ 
            & $2.32^{+0.14}_{-0.17}$ 
                    & $(8.14^{+2.02}_{-2.04}\times 10^{-3})$ 
            & 1.09 (454)  \\[2pt]  
 & Rev0045/301 & MOS1 (s7) & $9.27^{+3.17}_{-3.27}\times 10^{20}$ 
            & $0.74^{+0.02}_{-0.02}$ & $(8.93^{+1.22}_{-1.13})$  
            & $2.28^{+0.12}_{-0.14}$ 
                     & $(8.65^{+1.66}_{-1.71}\times 10^{-3})$ 
            & 1.02 (271) \\[2pt] 
 & Rev0066/101 & PN & $0.40^{+3.34}_{-0.40}\times 10^{20}$ 
            &  &   
            & $1.89^{+0.10}_{-0.06}$ 
                     & $1.85^{+0.14}_{-0.07}\times 10^{-4}$ 
            & 1.08 (54) \\[2pt] 
 & Rev0092/201 & PN & $6.19^{+0.55}_{-0.47}\times 10^{20}$ 
            & $0.22^{+0.01}_{-0.01}$ & $766.2^{+81.6}_{-62.0}$  
            & $2.05^{+0.02}_{-0.02}$ 
                     & $9.16^{+0.24}_{-0.20}\times 10^{-3}$ 
            & 1.15 (1000) \\[2pt] 
           &  & MOS1 (s7) & $3.07^{+0.91}_{-0.90}\times 10^{20}$ 
            & $0.26^{+0.01}_{-0.01}$ & $380.1^{+58.3}_{-50.1}$ 
            & $1.97^{+0.03}_{-0.04}$ 
                     & $1.11^{+0.05}_{-0.05}\times 10^{-2}$ 
            & 1.49 (376)  \\[2pt]             
\hline   
\end{tabular}  
\end{minipage} 
\end{table*}

\begin{table*}
\begin{minipage}{150mm}
\caption{Fits to the RGS data: 
   absorbed disk blackbody and power law }  
\label{mathmode} 
\centering 
\begin{tabular}{@{}lllccccclcc} 
\hline
 & Observation\footnote{We do not show fit parameters for 
observations without enough counts or without a reliable 
energy calibration. We assumed a metallicity of 0.4 times 
the solar value, i.e., Z = 0.008} 
       &  & $n_{\rm H}$\footnote{ 
                The Galactic line-of-sight absorption  
           ($= 3.2\times 10^{20}$~cm$^{-2}$) has been subtracted.}  
       & $T_{\rm in}$  & $A_{\rm disk}$ 
       & $\Gamma$\footnote{Fixed.}     & $A_{\rm pl}$ 
       & $\chi_\nu^2$(dof)      
  \\ \hline \hline 
&&&&&&&&\\
 & Rev0028/201 & RGS2 (s1) & $4.37^{+1.37}_{-0.35}\times 10^{20}$  
            & $0.55^{+0.02}_{-0.02}$ & $183.8^{+17.1}_{-15.6}$ 
            & 2.7 & $0.00^{+0.25}_{-0.00}\times 10^{-2}$  
            & 1.38(485)  \\[2pt]  
 & Rev0045/101 & RGS1 (s1) & $8.30^{+2.73}_{-2.85}\times 10^{20}$  
            & $0.63^{+0.07}_{-0.06}$ & $60.73^{+25.47}_{-22.18}$ 
            & 2.5 & $1.10^{+0.65}_{-0.58}\times 10^{-2}$  
            & 1.04(545)  \\[2pt]   
           & & RGS2 (s1) & $5.04^{+3.67}_{-1.07}\times 10^{20}$  
            & $0.49^{+0.03}_{-0.03}$ & $ 195.0^{+5.8}_{-38.4}$ 
            & 2.5 & $0.55^{+7.92}_{-0.55}\times 10^{-3}$  
            & 1.14(488)  \\[2pt]  
 & Rev0045/201 & RGS1 (s1) & $8.13^{+2.63}_{-2.78}\times 10^{20}$  
            & $0.67^{+0.08}_{-0.06}$ & $51.16^{+22.14}_{-19.01}$ 
            & 2.5 & $1.06^{+0.62}_{-0.56}\times 10^{-2}$  
            & 1.08(543)  \\[2pt]    
           & & RGS2 (s1) & $7.57^{+3.71}_{-2.92}\times 10^{20}$  
            & $0.50^{+0.03}_{-0.03}$ & $178.5^{+23.7}_{-40.9}$ 
            & 2.5 & $4.32^{+9.3}_{-4.32}\times 10^{-3}$  
            & 1.05(484)  \\[2pt]  
 & Rev0045/301 & RGS1 (s1) & $5.20^{+4.76}_{-1.38}\times 10^{20}$  
            & $0.53^{+0.05}_{-0.05}$ & $127.4^{+54.8}_{-41.7}$ 
            & 2.5 & $0.45^{+9.13}_{-0.45}\times 10^{-3}$  
            & 1.12(543)  \\[2pt]    
           & & RGS1 (s3) & $10.05^{+3.96}_{-4.28}\times 10^{20}$  
            & $0.70^{+0.18}_{-0.10}$ & $42.99^{+14.98}_{-24.12}$ 
            & 2.5 & $1.07^{+0.96}_{-0.81}\times 10^{-2}$  
            & 0.91(541)  \\[2pt]  
           & & RGS1 (s5) & $7.45^{+4.04}_{-2.24}\times 10^{20}$  
            & $0.53^{+0.06}_{-0.04}$ & $119.0^{+53.3}_{-38.5}$ 
            & 2.5 & $0.82^{+7.84}_{-0.82}\times 10^{-3}$  
            & 1.08(540)  \\[2pt]  
           & & RGS2 (s4) & $9.90^{+5.18}_{-2.73}\times 10^{20}$  
            & $0.45^{+0.03}_{-0.03}$ & $ 252.7^{+91.0}_{-74.7}$ 
            & 2.5 & $3.18^{+13.5}_{-3.18}\times 10^{-3}$  
            & 1.03(487)  \\[2pt]  
 & Rev0066/101 & RGS1 (s7) & $0.00^{+7.77}_{-0.00}\times 10^{20}$  
             & $0.07^{+0.03}_{-0.03}$ & $3311^{+4171}_{-3056}$ 
             & 1.9 & $2.77^{+0.20}_{-0.20}\times 10^{-4}$  
             & 1.11(44)  \\[2pt]  
 & Rev0092/101 & RGS1 (s2) & $5.85^{+1.03}_{-1.06}\times 10^{20}$  
            & $0.33^{+0.01}_{-0.02}$ & $211.2^{+29.6}_{-25.3}$ 
            & 2.05 & $0.93^{+0.26}_{-0.25}\times 10^{-2}$  
            & 1.27(538)  \\[2pt]    
           & & RGS2 (s2) & $8.01^{+0.95}_{-1.12}\times 10^{20}$  
            & $0.27^{+0.02}_{-0.02}$ & $478.1^{+103.8}_{-81.9}$ 
            & 2.05 & $1.36^{+0.28}_{-0.14}\times 10^{-2}$  
            & 1.25(485)  \\[2pt]   
 & Rev0092/201 & RGS1 (s4) & $6.40^{+1.35}_{-1.45}\times 10^{20}$  
            & $0.29^{+0.03}_{-0.03}$ & $252.0^{+83.9}_{-52.9}$ 
            & 2.05 & $1.06^{+0.28}_{-0.31}\times 10^{-2}$  
            & 1.08(537)  \\[2pt]    
           & & RGS2 (s5) & $6.84^{+1.19}_{-1.28}\times 10^{20}$  
            & $0.27^{+0.03}_{-0.03}$ & $500.0^{+65.1}_{-52.7}$ 
            & 2.05 & $0.99^{+0.09}_{-0.10}\times 10^{-2}$  
            & 1.15(485)  \\[2pt]   
\hline   
\end{tabular}  
\end{minipage} 
\end{table*}

\begin{figure} 
\begin{center}
\setlength{\unitlength}{1cm}
\begin{picture}(6.6,5.8)
\put(0,0){\includegraphics{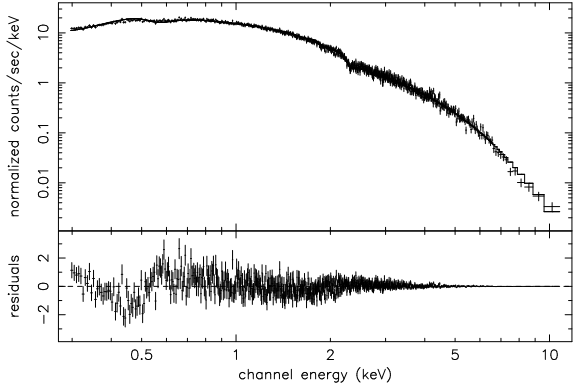}}
\end{picture}
\caption{The top panel shows 
   the data and fit EPIC-PN spectra 
   of the Rev0028/201 observation (2000 February 2).    
   An absorbed diskbb$+$powerlaw model is used in the fit 
   (see Table 3). 
   The residuals are shown in the bottom panel.} 
\label{fig2} 
\end{center} 
\end{figure}

\begin{figure} 
\begin{center}
\setlength{\unitlength}{1cm}
\begin{picture}(6.6,5.8)
\put(0,0){\includegraphics{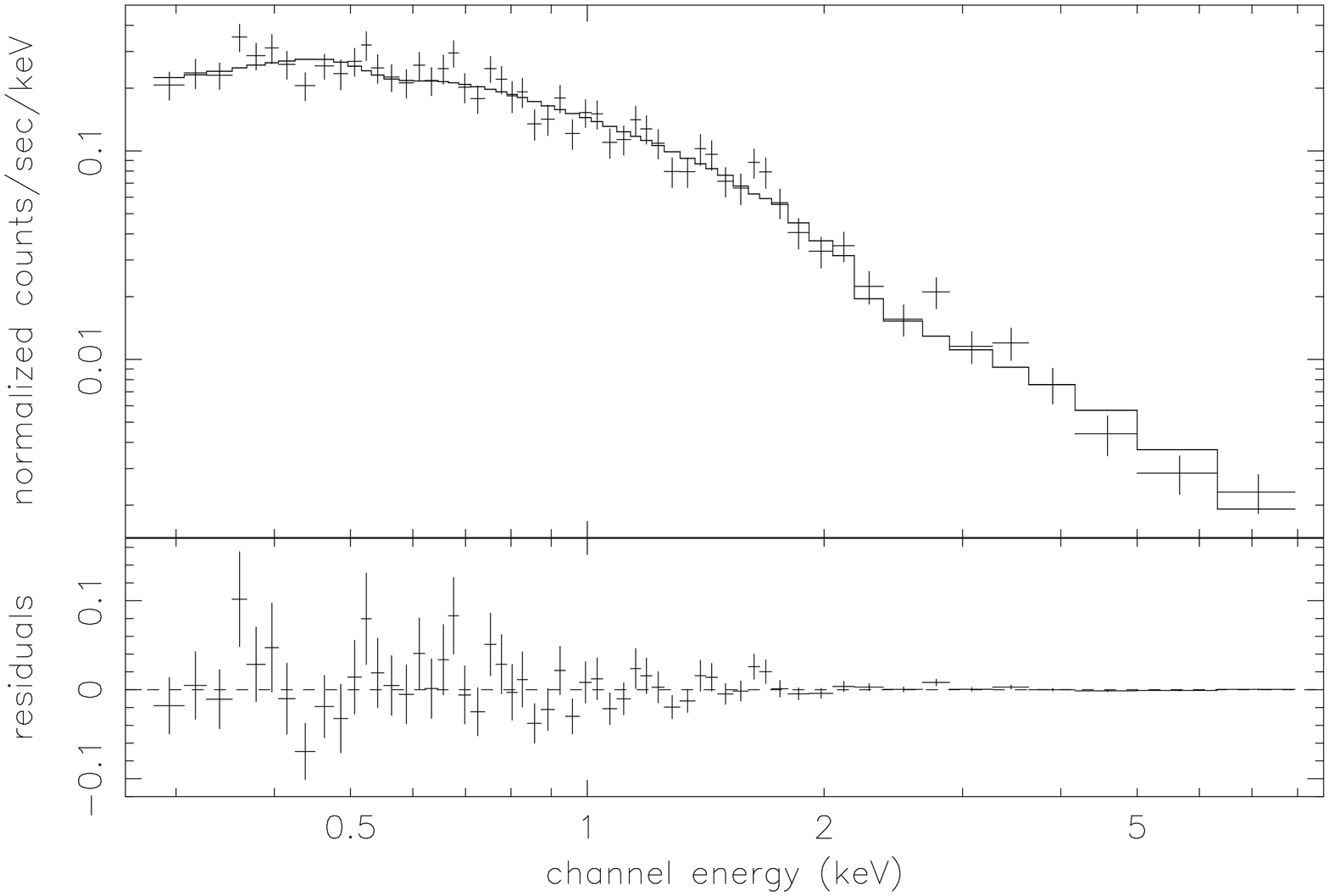}}
\end{picture}
\caption{Same as Fig.~2 for the Rev0066/101 EPIC-PN observation 
         (2000 April 19).}
\label{fig3} 
\end{center} 
\end{figure}

\begin{figure} 
\begin{center}
\setlength{\unitlength}{1cm}
\begin{picture}(6.6,5.8)
\put(0,0){\includegraphics{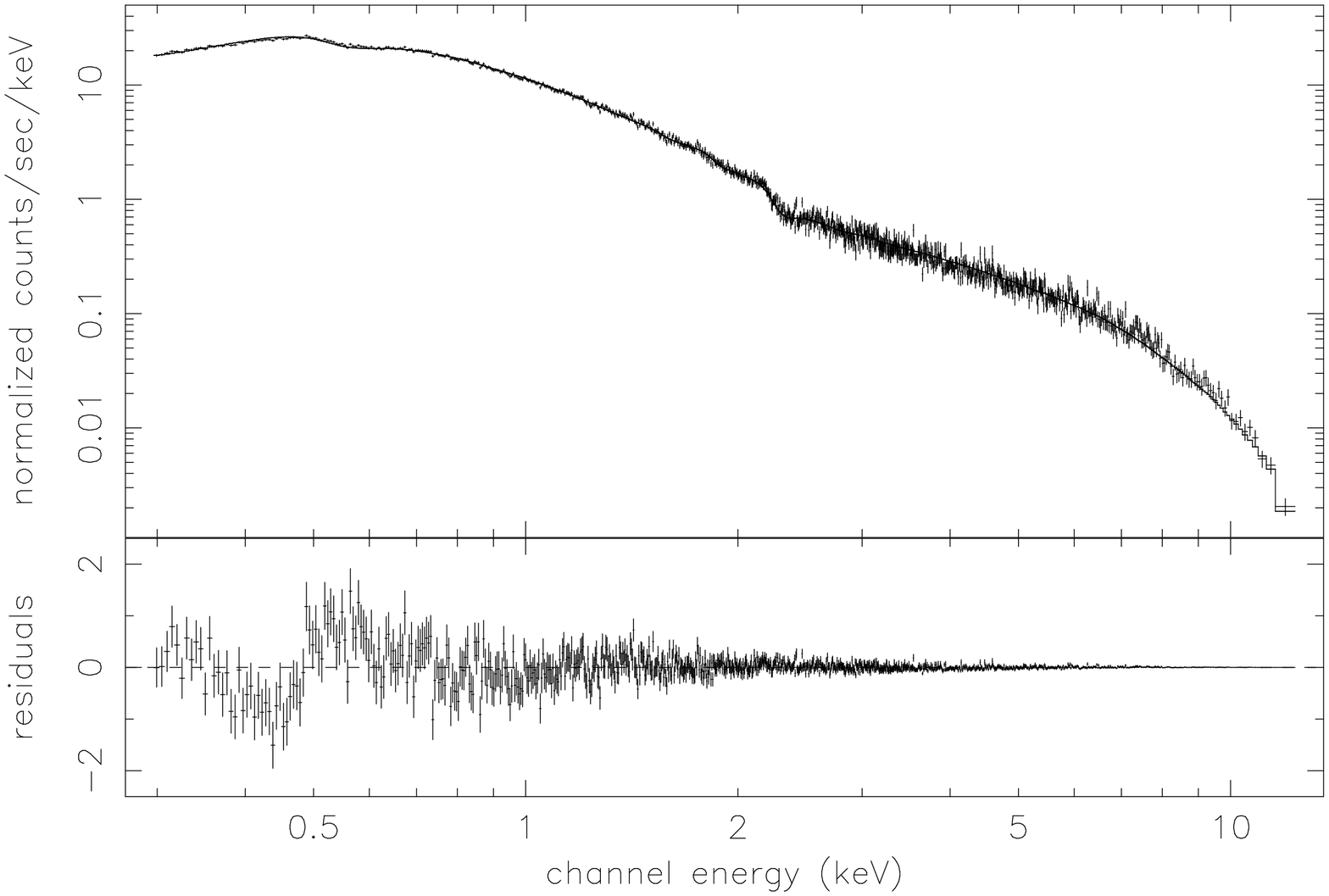}}
\end{picture}
\caption{Same as Fig.~2 for the Rev0092/201 EPIC-PN observation
         (2000 June 10).}
\label{fig3} 
\end{center} 
\end{figure}    

\subsection{RGS data}  

We consider the RGS data 
   to search for possible emission lines, 
   and to constrain the value of the column density. 
The RGS spectra are fitted 
  with the same model used in the analysis of the PN data.  
As the RGS spectra cover only the 0.35$-$2.5~keV energy range, 
  the thermal and the power-law components cannot be constrained  
  simultaneously. 
We therefore fix $\Gamma$ 
  to the values determined from the EPIC fits 
  (with a correction to the value of $\Gamma$ from Rev0045 
  to take the pile-up into account).   
Because of the low count rate, 
  the data in the Rev0066 observations   
  are binned in groups of 100 channels each.  
The error-weighted means of the fit values of $T_{\rm in}$  
  are 0.52 keV and 0.29 keV 
  for observations before and after Rev0066 respectively.   

We do not find strong evidence of emission lines. 
The column density $n_{\rm H}$ was generally below $10^{21}$~cm$^{-2}$,  
  except for Rev0066, 
  where $n_{\rm H}$ is not well constrained.    
After the line-of-sight column density to the LMC is subtracted, 
  the value of the column density for the observations 
  before Rev0066 is $(5.50\pm0.67) \times 10^{20}$~cm$^{-2}$, 
  and the value after Rev0066 is $(6.83\pm0.58) \times 10^{20}$~cm$^{-2}$.   
  (These values are smaller than those 
  obtained from the fits to the EPIC data.)  
Given that the error may be dominated by systematic effects,   
  the error obtained in the fits may not truly represent 
  the statistical weights of the fit parameters.  
Apart from the conclusion that 
  the spectra do not show strong column absorption, 
  we are unable to determine whether the variations in $n_{\rm H}$   
  are correlated with the X-ray luminosity.

\section{Discussion}   
         
\subsection{Soft and hard states}     

Previous {\xte}/PCA observations (Wilms et al.\ 2000) 
  have shown that 
  when the source was bright 
  (with {\xte}/ASM count rates $\simgt 3$~ct~s$^{-1}$), 
  the disk blackbody component was prominent, and 
  its fit temperature $T_{\rm in} \approx 1.0$~keV. 
There was also a power-law tail with $\Gamma \approx 4$ in the spectra. 
The temperature $T_{\rm in}$ appeared to decrease 
  when the X-ray luminosity decreased,  
  while the normalisation parameter $A_{\rm disk}$ 
  remained approximately constant.   
When the {\xte}/ASM count rate dropped below $\approx 0.6$~ct~s$^{-1}$,  
  $T_{\rm in}$ was reduced to $\approx 0.7$~keV. 
The transition was accompanied by the hardening of the power-law component.  
The photon index $\Gamma$ became $\sim 2.0$--$3.0$. 
When the {\xte}/ASM count rate decreased 
  below the 0.3-ct~s$^{-1}$ level, 
  the disk blackbody component was not detected, 
  and the spectrum was a power law with $\Gamma = 1.8$. 

The soft-to-hard transition was also seen in our data.  
The spectra obtained in the Rev0028--Rev0045 observations 
  show a disk blackbody component with $T_{\rm in} \sim 0.6$--$1.0$, 
  and a power-law component with $\Gamma \approx 2.5$.  
The {\xte}/ASM count rate was below $2$~ct~s$^{-1}$ 
  during the observations.  
The spectral properties are similar to 
  those observed previously 
  when the system had similar {\xte}/ASM count rates. 

The spectrum obtained in the Rev0066/101 observations 
  is dominated by a power law with $\Gamma = 1.9 \pm 0.1$.  
The photon index is consistent with that observed 
  in the previous hard states  
  (Wilms et al.\ 2000), while a lower value of $1.60\pm0.04$
  was obtained by Boyd et al.\ (2000) on 2000 May 7.   
The $\chi^2$ of the fit which 
  includes a thermal disk black-body component 
  (best fit $T_{\rm in} = 0.14$~keV)
  is 58.486 for 52 degrees of freedom.  
If the thermal component is not included, 
  then we obtain $\chi^2 = 58.493$ for 54 degrees of freedom. 
The thermal component is therefore insignificant. 

The system was at the transition 
  from the hard state back to the soft state 
  during the Rev0092 observations (see Figure~1). 
The power-law component had steepened, 
  with $\Gamma = 2.05 \pm 0.02$ (for the PN data).  
The disk blackbody component reappeared, with a temperature 
  $T_{\rm in}\approx 0.2$~keV, significantly lower than 
  that before Rev0066. 
 
In summary, between 2000 February and June  
  \lmcx3 underwent a transition 
  from a soft to a hard state, 
  and then in the process of returning to the soft state.    
During the soft-to-hard transition 
  the fit temperature 
  of the disk blackbody component decreased, 
  and the power law component became harder. 
As the system started to return to the soft state, 
  the disk blackbody component became more prominent.  
  The power law appeared to be steeper than that 
  obtained from the RXTE/PCA observations near the middle of the 
  faint state, yet the disk blackbody temperature was still well below 
  the values of the previous soft state.

\subsection{Mode of mass transfer}    
 
The RGS data 
  constrain the value of $n_{\rm H}$ within the system 
  to be $\simlt 10^{21}$~cm$^{-2}$, 
  in contrast to the larger intrinsic column density    
  expected for a companion with a strong stellar wind. 
The non-detection of obvious emission lines 
  in the RGS spectra also indicates 
  the absence of wind matter ejected in previous epochs 
  (cf. the P Cygni lines seen in Cir X-1, Brandt \& Schulz 2000). 
Thus, the {\xmm} spectral data 
  do not support the mass-transfer scenario for \lmcx3 
  in which the black hole accretes matter 
  mainly from a strong stellar wind from a massive companion. 
The high luminosity observed in the X-ray bands therefore 
  requires the companion to overflow its Roche lobe.  

A similar conclusion is also obtained independently 
  from an analysis of the optical/UV properties of the system 
  (Soria et al.\ 2001).  
The OM data obtained in the Rev0066 
  suggest that the companion is a B5 subgiant 
  instead of a B3 main-sequence star. 
Roughly 3\% of the X-rays would be intercepted  
  by the companion, so that 
  the rate at which energy is deposited into its atmosphere 
  can be $\simgt 5\times 10^{36}$~erg s$^{-1}$ in the soft state. 
This rate is larger than 
  the intrinsic luminosity of the companion.   
If the companion in \lmcx3 is indeed a subgiant star 
  its tenuous envelope is susceptible to irradiation heating.     
The soft-to-hard transitions 
  seen in the {\xte} observations in 1997/1998 
  and the {\xmm} observations in 2000 may be caused by 
  variations in the rate of mass overflow 
  from the Roche lobe of the subgiant companion.  

Wilms et al.\ (2000)'s interpretation of the decreases in the 
  RXTE/ASM count rate as evidence for transitions from soft to hard 
  state is consistent with our data. We further propose that the 
  decrease in the X-ray luminosity is caused by the decrease 
  in the fraction of the Roche lobe filled by the companion star. When the 
  companion is detached from its critical Roche surface, mass 
  transfer will be dominated by a focused wind.

It is worth noting that the three known high-mass BHCs   
  are all persistent X-ray sources 
  which show preferential X-ray spectral states. 
While Cyg X-1 tends to be in the hard state, 
  LMC X-1 and \lmcx3 are more often found in the soft state.  
Recent studies  
  (e.g.\ Igumenshchev, Illarionov \& Abramowicz 1999;  
  Beloborodov \& Illarionov 2000)    
  have shown that 
  accretion of matter with low angular momentum 
  will give rise to hard X-rays instead of soft X-rays.  
The relative angular momentum of the accreting matter 
  is smaller for wind accretion than for Roche-lobe overflow. 
Cyg X-1 has a 33-$M_{\odot}$ O-type companion  
  (Giles \& Bolton 1986), which has a strong stellar wind. 
The companion stars in LMC X-1 and \lmcx3 
  are less massive ($\simlt 10$~M$_\odot$) B stars,  
  whose stellar wind is much weaker.    
Therefore, we suggest a unified scenario 
  which relates the mode of mass transfer (wind or Roche-lobe overflow) 
  to the spectral state preferentially observed 
  in these three high-mass BHC binaries (hard or soft respectively).  
The subgiant companion of \lmcx3 
  may occasionally underfill its Roche lobe  
  because of feedback irradiative processes 
  or instabilities in its envelope.   
This leads to the residual accretion of the (focused) wind matter 
  which has relatively low angular momentum.  

\section{Summary}   

The BHC \lmcx3 was observed in 2000 February--June,  
   with the {\xmm} EPIC and RGS, 
   throughout a soft-hard transition. 
The system was apparently in the process 
   of returning to the soft state in 2000 June. 
The hard-state spectra are dominated by a power-law component             
   with a photon index $\Gamma \approx 1.9$. 
The soft-state spectra consist of a thermal component  
   with an inner-disk temperature $T_{\rm in}$ 
   of $\approx 0.9$~keV   
   and a power-law tail with $\Gamma \approx 2.5$--$2.7$.   
The line-of-sight absorption deduced from the EPIC and RGS data   
   is $n_{\rm H} \simlt 10^{21}$~cm$^{-2}$. 
Our observations therefore do not support the wind accretion 
   model for this system in the soft state.
The transition from the soft to the hard state     
   appears to be a smooth process 
   associated with the changes in the mass-transfer rate.     
 
\begin{acknowledgements}  
  This work is based on observations obtained with {\xmm}, 
     an ESA science mission with instruments and contributions 
     directly funded by ESA member states and the USA (NASA). 
  We thank Keith Mason for his comments.
  KW acknowledges a PPARC visiting fellowship. 

\end{acknowledgements}

\end{document}